\documentclass[prl,twocolumn,aps]{revtex4}

\begin{document}
\textbf{Comment on "Absence of the Mott Glass Phase in 1D Disordered
  Fermionic Systems''}

In Refs.\cite{orignac_mg_short,giamarchi_mottglass_long} we predicted
a novel phase for interacting fermions (or bosons)
in presence of quenched disorder, intermediate between the
Mott Insulator (incompressible with a gap in the optical conductivity)
and the Anderson glass (compressible with a pseudo-gap in the
optical conductivity). In this phase, called the Mott Glass, the system
is incompressible but exhibits only a pseudo gap in the optical conductivity).
This prediction was based on semiclassical arguments, excitonic arguments,
and supported by a variational replica calculation and a functional RG calculation.
It required finite extent repulsive interactions, at least nearest neighbors.

In Ref.~\cite{nattermann_mottglass_absence}, it is claimed that such an intermediate
phase cannot exist. The analysis is based on an extension of the semiclassical
treatment \cite{nattermann_temperature_luttinger} of the Anderson insulator
to the case of a Mott insulator with disorder. In Ref.~\cite{nattermann_mottglass_absence}
the energy of kinks (i.e. topological excitations) are computed.
Ref.~\cite{nattermann_mottglass_absence} argues, based on their Eq. (4),
that a non-zero optical conductivity $\sigma(\omega)$ requires a
zero-energy gap to kink formation in the system and thus a non-zero
compressibility.

Although Eq. (4) is technically correct, we want to point out in
the present comment
that the subsequent argumentation in insufficient to rule out the existence of the
Mott glass.
Indeed as was already shown in \cite{giamarchi_mottglass_long} (see
Fig. 5), in the phase representation, the Mott glass phase is characterized by
a Hessian matrix with a positive spectrum extending down to zero and
a kink gap remaining strictly positive.  As shown in
Sec. III A of \cite{giamarchi_mottglass_long}, the incompressibility stems
from the  strictly positive kink gap,
while the source of the non-vanishing conductivity
is the spectrum of the Hessian matrix extending down to zero. In other
words, the non-vanishing frequency dependent conductivity
in the Mott glass state
is caused by \emph{non-topological} low-energy excitations. A sketch
of  such non-topological excitations present in the Mott glass
can be seen on Fig.~5 of \cite{giamarchi_mottglass_long} which shows
the phase field remaining bounded within an interval of magnitude strictly less
that $\pi$, but exhibiting multiple extrema within this interval.  As
Eq. (4) in \cite{nattermann_mottglass_absence} only contains contribution from the
topological excitations of the field,  these non-topological
excitations are overlooked in  the considerations of Ref.~\cite{nattermann_mottglass_absence}, so that only the Mott and the Anderson insulating state
can be recovered.

Examples of contributions of non-topological excitations to optical
conductivity besides kink contributions are known to occur  in several related
situations. For instance in the  non-disordered sine-Gordon model,
which corresponds to the pure Mott insulator,
non-topological excitations (breathers) with lower energy than the
kinks (solitons) can exist. Applying the reasoning of
\cite{nattermann_mottglass_absence} to the sine-Gordon model would lead to the conclusion that the
conductivity gap in this model is always larger than the soliton gap, which is clearly not the case for sufficiently small parameter
$K$ in the sine-Gordon model \cite{rajaraman_instanton,essler_mott_excitons1d}.
For the non-commensurate disordered case, non topological excitations
do also lead to a non-zero absorption at finite frequency, and thus a non-zero optical
conductivity, even if kinks are totally excluded such as
in the extreme classical limit \cite{fogler_cdw}.

For the commensurate disordered case, as pointed out in Ref.~\cite{orignac_mg_short,giamarchi_mottglass_long}
the non-topological excitations are thus at the root of the existence of the Mott glass phase.
In a direct fermionic language they correspond to particle-hole bound pairs, similar to excitons,
while kinks corresponds to independent particles. An explicit example
is given in Section IV B of Ref.\cite{giamarchi_mottglass_long}.
While excitons, being
neutral particles, do not contribute to dc transport they can and do give a contribution
to the optical conductivity.

Besides the derivations given in Ref.~\cite{orignac_mg_short,giamarchi_mottglass_long},
it is also interesting to point out that a real space renormalization
procedure on disordered commensurate bosons,
controlled in the limit of strong disorder, also leads to a similar
conclusion to ours on the existence of a gapless incompressible phase
\cite{altman_mottglass_rsrg}.
There it is shown that this phase can be viewed as a chain
of non-interacting finite superfluid grains; as the thermodynamic limit
is taken, the lowest charging gap (i.e., kink energy) vanishes as
1/log(L), due to essentially a single anomalously large grain. Adding
particles to this grain hardly affects the {\it density}, and thus the
compressibility is only: $\kappa~\log(L)/L\rightarrow 0$ as $L\rightarrow
\infty$, disproving the statement in Ref.~\cite{nattermann_mottglass_absence} (page 3
under table 1) that zero compressibility implies a finite gap.

\noindent Pierre Le Doussal$^1$, Thierry Giamarchi$^2$, and Edmond Orignac$^3$ \\
${}^1$LPTENS, CNRS UMR8549, 24 Rue Lhomond, 75231 Paris Cedex 05
France \\
${}^2$ DPMC-MaNEP, Universite de Geneve, 24 Quai Ernest Ansermet, CH1211
Geneve, Switzerland \\
${}^3$ LPENSL, CNRS UMR5672, 46 All\'ee d'Italie, 69364 Lyon Cedex 07,
France\\

\end{document}